\documentclass[a4paper,fleqn,usenatbib]{mnras}

\usepackage[T1]{fontenc}
\usepackage{ae,aecompl}

\usepackage{graphicx}    
\usepackage{amsmath}    
\usepackage{amssymb}    
\usepackage{subcaption}
\usepackage{booktabs}
\usepackage[flushleft]{threeparttable}

\newcommand{\mps}{m s$^{-1}$}    

\title[Robustness of Keplerian signals to the removal of active and telluric features]{Exploring the robustness of Keplerian signals to the removal of active and telluric features}

\author[M. Lisogorskyi et al.]{
M. Lisogorskyi\thanks{E-mail: m.lisogorskyi@herts.ac.uk}$^{1}$,
H. R. A. Jones$^{1}$,
F. Feng$^{2}$,
R. P. Butler$^{2}$
and S. Vogt$^{3}$
\\
$^{1}$Centre for Astrophysics Research, University of Hertfordshire, College Lane, AL10 9AB, Hatfield, UK\\
$^{2}$Earth and Planets Laboratory, Carnegie Institution of Washington, Washington, DC 20015, USA\\
$^{3}$UCO/Lick Observatory, University of California, Santa Cruz, CA 95064, USA
}

\date{Accepted 2020 October 6. Received 2020 September 29; in original form 2020 June 15}

\pubyear{2020}

\begin{document}
\label{firstpage}
\pagerange{\pageref{firstpage}--\pageref{lastpage}}
\maketitle

\begin{abstract}
We examine the influence of activity- and telluric-induced radial velocity signals on high resolution spectra taken with an iodine absorption cell.
We exclude 2 Angstrom spectral chunks containing active and telluric lines based on the well characterised K1V star Alpha Centauri B and illustrate the method on Epsilon Eridani--an active K2V star with a long period low amplitude planetary signal.
After removal of the activity- and telluric-sensitive parts of the spectrum from the radial velocity calculation, the significance of the planetary signal is increased and the stellar rotation signal disappears.
In order to assess the robustness of the procedure, we perform Monte Carlo simulations based on removing random chunks of the spectrum.
Simulations confirm that the removal of lines impacted by activity and tellurics provides a method for checking the robustness of a given Keplerian signal.
We also test the approach on HD 40979 which is an active F8V star with a large amplitude planetary signal.
Our Monte Carlo simulations reveal that the significance of the Keplerian signal in the F star is much more sensitive to wavelength.
Unlike the K star the removal of active lines from the F star greatly reduces the radial velocity precision.
In this case, our removal of a K star active lines from an F star does not a provide a simple useful diagnostic because it has far less radial velocity information and heavily relies on the strong active lines.
\end{abstract}


\section{Introduction}
\label{sec:introduction}

Using iodine cells as wavelength calibrators allows achieving radial velocity (RV) precision of 1--3 \mps{}\citep{1996pasp..108..500b, 2020PASP..132a4503X} and it has proven to be a very effective technique for exoplanet detection \citep[e.g.][]{1999ApJ...526..916B, 2004ApJ...617..580B, 2017AJ....153..208B}.
Even with this instrumental precision, stellar activity can be the major source of noise.
The usual way of distinguishing the planetary signals from activity is to compare or decorrelate various activity indicators like Ca \textsc{II} H\&K emission \citep[e.g.][]{1991ApJS...76..383D}, or cross-correlation function (CCF) bisectors \citep[e.g.][]{Gray_2005}.

Spectral indices trace activity variations at a specific wavelength in the spectrum and originate from a particular physical location in the atmosphere of the star.
Chromospheric emission is good at predicting radial velocity variations due to activity, but almost all radial velocity information comes from photospheric absorption lines.
CCF bisectors, on the other hand, contain information on activity variations of the whole spectrum, but it is averaged out.
Spectral lines originate in different pressure and temperature environments, different parts of the star and are affected by the magnetic fields to a different degree.
There have been many studies looking for new spectral indices beyond the $S$-index and $R_{HK}$ to trace the activity across the spectrum \citep[e.g.][]{2017MNRAS.468L..16T, 2018AJ....156..180W, 2019MNRAS.485.4804L} or to correct for line variation \citep{2018A&A...620A..47D, 2020A&A...633A..76C}.
Another approach to the problem is to test signal's wavelength dependence.
For instance, \citet{2020A&A...638A...5C} used multi-band spectroscopic observations from GIANO-B and HARPS-N \citep{2017EPJP..132..364C} to investigate a signal that showed wavelength dependence incompatible with a Keplerian orbit.
\citet{2017AJ....154..135F} and \citet{2017A&A...605A.103F} used radial velocities from 72 \'echelle orders of HARPS to get a robust model of wavelength-dependent noise.

In this work we use a list of active and telluric lines \citep[see Sections \ref{sec:actlines} and \ref{sec:tellines} or][]{2019MNRAS.485.4804L} and attempt to remove them from radial velocity calculation using iodine data.
The iodine processing technique provides separate radial velocity measurements for each of 718, $\sim$2\AA-wide wavelength intervals.
Given that individual lines identified as active are typically less than 0.5\AA{}-wide \citep[based on][]{2019MNRAS.485.4804L} this provides a ``natural'' route as part of the data processing to eliminate active lines from the calculation of radial velocity without the need to remove such large spectral regions as has been done in previous work.
The spectral lines in our selection show different shape variations, which might manifest in different signals and magnitude of the effect on the radial velocity measurements, but in this work we treat them equally.

As a test, we choose an active star ($\epsilon$ Eri) with a rotational and a planetary signal in the radial velocity data (Section \ref{sec:results:hd22049}).
In addition, we investigate usefulness of Monte Carlo methods in the analysis of a signal's wavelength dependence.
Bootstrap methods can be, but scarcely are, used to extract correct uncertainties on a planetary fit \citep{BALUEV201318}.
Even fewer studies do rigorous analysis of wavelength dependence of a signal, which would ensure robustness of a planetary detection \citep[e.g.][]{2017AJ....154..135F,2017A&A...605A.103F,2018AJ....155..192T, 2020A&A...638A...5C}.
We quantify Keplerian signal robustness using this method with our active and telluric lines lists and further test on another active star HD 40979 with a different spectral type (Section \ref{sec:results:hd40979}).

\section{Data and methods}

\begin{figure*}
    \centering
    \begin{subfigure}{\textwidth}
        \centering
        \includegraphics[width=\textwidth]{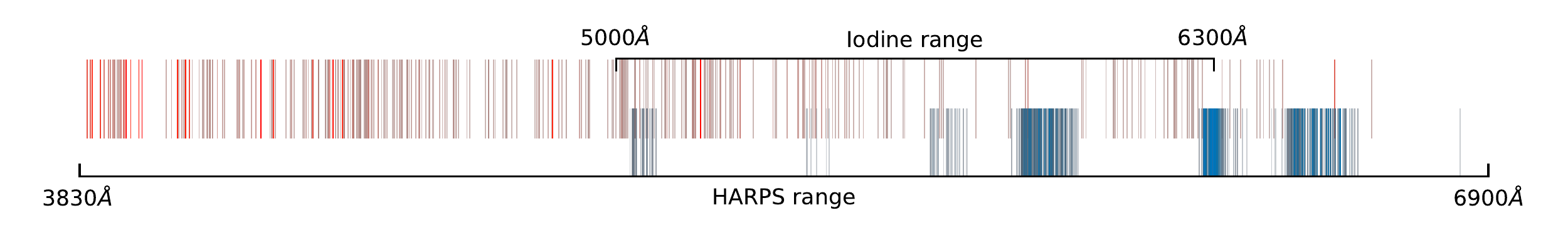}
        \label{fig:smth}
    \end{subfigure}
    \caption{
        A qualitative comparison of the HARPS wavelength range, used to identify the active and telluric lines in $\alpha$ Cen B, and the iodine range, used in this work.
        The active lines are shown in red as vertical lines and colour coded from the lowest variation amplitude (gray) to the highest (red).
        The telluric lines are shown in blue as vertical lines and colour coded from the lowest relative depth (gray) to the highest (blue).
    }
\label{fig:lines}
\end{figure*}

We assess the impact on radial velocity signals of removing spectral regions to quantify both impact of activity and Keplerian signal robustness.
The stars were chosen based on being active stars with planetary signals.
Figure \ref{fig:lines} shows the location of the lines in the spectrum.

\subsection{Active line selection}
\label{sec:actlines}

The active lines were selected using HARPS observations of $\alpha$ Cen B during the 2010 observing season.
The data set contains 2277 high resolution and high signal-to-noise ratio spectra, providing good phase coverage of the rotation of the star.
During that period the star was at the peak of its activity cycle and showed higher than normal amplitude activity variations within the range of log$R_{HK}$ = -4.82 to -4.95 \citep{2012Natur.491..207D}.
During this 2010 dataset the star shows high amplitude activity variation clearly modulated by the rotation of the star.
The rotational signal is visible both in log$R_{HK}$ and RV, which provides an activity variation baseline.

The quality of this dataset makes it very useful for activity studies.
\citet{2017MNRAS.468L..16T} used a similar version of this archive data to compare ``active'' spectra to ``inactive'' and study variations of iron lines.
\citet{2018AJ....156..180W} supplemented the $\alpha$ Cen B data with HARPS data from $\epsilon$ Eri to present a new semi-automatic method for finding activity sensitive lines, resulting in a list of 40 new activity indicators.
\citet{2018A&A...620A..47D} and \citet{2020A&A...633A..76C}, using a similar version of the HARPS $\alpha$ Cen B archive data set, demonstrated a method of mitigating stellar activity noise by measuring radial velocities from individual spectral lines.

We identify the activity indicators using a method similar to the one used by \citet{2017MNRAS.468L..16T}, extended for the whole HARPS spectral range.
Every spectrum is divided by a low activity template (10 spectra with the lowest value of $S$-index stacked) and inspected for pseudo-emission features.
In contrast with the method from \citet{2018A&A...620A..47D} and \citet{2020A&A...633A..76C}, which characterises line variability with a single number -- radial velocity.
The \citet{2017MNRAS.468L..16T} methodology potentially provides more information on different line variation types -- shift, width, core flux, wings, blended lines etc.
All the features we have selected correlate with $S$-index and radial velocity in the data set considered, either in flux or asymmetry.
Even the features that seem to produce a symmetric shape variation impact the radial velocity via changing the width of the CCF and the feature producing it moving across a stellar disk \citep[as shown in][]{2017MNRAS.468L..16T, 2019MNRAS.485.4804L}.

This method is, however, biased towards deeper lines -- the ones that carry the most amount of Doppler information in the spectra.
Most lines in our selection ($\sim72$ per cent) are deeper than 0.8 (relative to a normalised continuum of 1) and removing many deep features from the spectra might have a big impact on the radial velocity precision, depending on the number of lines available.
This is discussed further in Section \ref{sec:results}.

The full list of lines used in this work is published in \citet{2019MNRAS.485.4804L}, and we treat it as a full list of active lines that definitely vary with activity.
It is worth noting that these lines' variability is most prominent during the high activity epoch of $\alpha$ Cen B and correlate with rotation of the star, which means they could largely come from specific features on the surface of the star.
In addition, the list is derived using K1V star spectra and is most useful for close spectral types.
While we expect these lines to be present to some extent in slightly hotter stars, cooler stars like M dwarfs will have a different set of variable spectral lines.

\subsection{Telluric line selection}
\label{sec:tellines}

We derive an empirical list of telluric lines using a similar method.
In \citet{2019MNRAS.485.4804L} we quantified telluric contamination of the spectra by measuring the equivalent width of a water line at 6543.9\AA{} as it does not overlap with any strong  stellar features over the course of a year.
We computed a telluric-free template of the stellar spectrum by averaging the least contaminated spectra.
In addition to having virtually no telluric lines present in these spectra, these observations were taken during different positions of the Earth around the Sun, which ensures that all the remaining features are averaged out.
To derive a telluric template, we choose a spectrum with the most prominent water line and divide it by the telluric-free template.
The resulting relative spectrum contains only telluric features and we extract all lines which appear to be reliably above the noise (a relative depth of 2 per cent).
The total number of features extracted is 460.
These are listed in the Appendix with corresponding measured rest-frame wavelengths and relative depths of the lines.

It is worth noting that these features are the strongest telluric lines in the observations considered and might not include the micro-tellurics that were also shown to have a non-negligible effect on radial velocity measurements \citep[e.g.][]{2014A&A...568A..35C}.
While this list of lines is not representative of all tellurics that can be observed from all observatories, it contains a large number of deep atmospheric lines and the data used to compile it spans almost a decade and wide range of humidity \citep[see][]{2019MNRAS.485.4804L}.

\subsection{Iodine data}
\label{sec:iodine}

Iodine is used as a wavelength calibrator for precise Doppler spectroscopy, achieving precision of 1--3 \mps{}\citep{1996pasp..108..500b}.
The light from a star is passed through an iodine cell before entering a spectrograph, imprinting narrow lines on top of the stellar spectrum.

The data reduction for iodine data provides radial velocity measurements across the spectrum based on individual 2\AA{} chunks.
As we want to exclude particular spectral lines from the radial velocity calculation, these can be identified in particular 2\AA{} chunks and removed from the calculation of radial velocity.
This makes iodine absorption cell data particularly easy to use for this technique as radial velocities for each chunk are produced as part of the standard pipeline processing.
After removing cosmic rays and other spurious signals, the remaining measurements are weighted and combined to get a ``true'' radial velocity.
The data provides wavelength coverage from approximately 5000\AA{} to 6300\AA{} \citep{1992PASP..104..270M}, which includes less than half the spectral lines in our list derived from the broader wavelength window used by HARPS.
The hypothesis is that if the chunks that contain activity sensitive lines are removed, it would reduce the significance of the stellar rotation signal and relatively increase the significance of a planetary signal.

The number of chunks removed depends on a number of ``active'' spectral lines to be removed and the absolute radial velocity of the star.
In case of HD 22049 we remove 137 out of 718 chunks with active lines and 121 with telluric lines, which increases the estimated uncertainties of the measurements (see Section \ref{sec:results:impact}), but removes unwanted signals.
Figure \ref{fig:lines} is a qualitative plot, showing locations of the lines in the spectrum and compares the iodine range to HARPS range.

\subsection{Monte Carlo method}
\label{sec:mc-intro}

To make sure that our line selection actually impacts the radial velocity measurements in a non-random way, we perform a Monte Carlo test.
On the basis that activity correlates with rotation, if we remove the same number of randomly selected chunks from the spectrum, the rotation signature should be stronger than when we remove the chunks we identified to contain active lines.

We use \textsc{Agatha} software for periodic analysis and compute Bayes Factor periodograms \citep[BFPs,][]{feng17agatha} using a Keplerian periodic model.
Bayesian factor ($ln$ BF) is used as a measure of signal significance as it represents a likelihood ratio of two competing hypotheses -- periodic model and the noise model.
If $ln$ BF is larger than 5, the periodic model is favoured over the noise model and the signal at this period is considered significant.
\textsc{Agatha} allows optimization of the noise model for a particular dataset, which we used in this work.
Red noise model with a single moving average (MA) component was determined to fit the $\epsilon$ Eri data set best, and a white noise model was used for HD 40979.

As some signals (e.g. rotation signal) have multiple peaks of varying height, we measure the significance of the highest peak within a small range around the signal (e.g. within 0.5 day of the 11.68 rotation period of $\epsilon$ Eridani).
To account for the limited resolution of the computed periodogram, we fit a parabolic function to the top three points of a peak and take its vertex.
We find that all the values increase only slightly, which means that the resolution is appropriate for the peaks we're investigating.

The Monte Carlo results including the RV data sets analysed in this work and the corresponding BFPs are published on \textsc{github}\footnote{\url{https://github.com/timberhill/iodine-monte-carlo}}.

\section{Results}
\label{sec:results}

\subsection{HD 22049}
\label{sec:results:hd22049}




HD 22049 ($\epsilon$ Eridani) is a K2V star \citep[e.g.][]{1989ApJS...71..245K}, with a $2691.8 \pm 25.6$ days period planet around it \citep{2019AJ....157...33M} with a signal amplitude of $11.49\pm0.66$ \mps.
It is quite an active star with log $R'_{HK}=-4.455$ \citep{2007ApJ...669.1167B}, more active than $\alpha$ Cen B with log $R'_{HK}=-4.923$ \citep[e.g.][]{2006ESASP.624E.111M} though as indicated earlier it should be recognised that a star does not have a fixed value of log $R'_{HK}$ and it is established that $\alpha$ Cen B varies between at least -4.8 and -5.0 \citep{2020A&A...633A..76C}.
This star is a good candidate to test due to its high activity, available data and a spectral type close to $\alpha$ Cen B, so we expect most of the lines we selected might be sensitive to activity in a similar way.
The data set has 104 observations over 8 years from HIRES instrument on Keck telescope, covering a little more than one period of the planet.

\begin{figure*}
    \centering
    \begin{subfigure}{.99\textwidth}
        \centering
        \includegraphics[width=\textwidth]{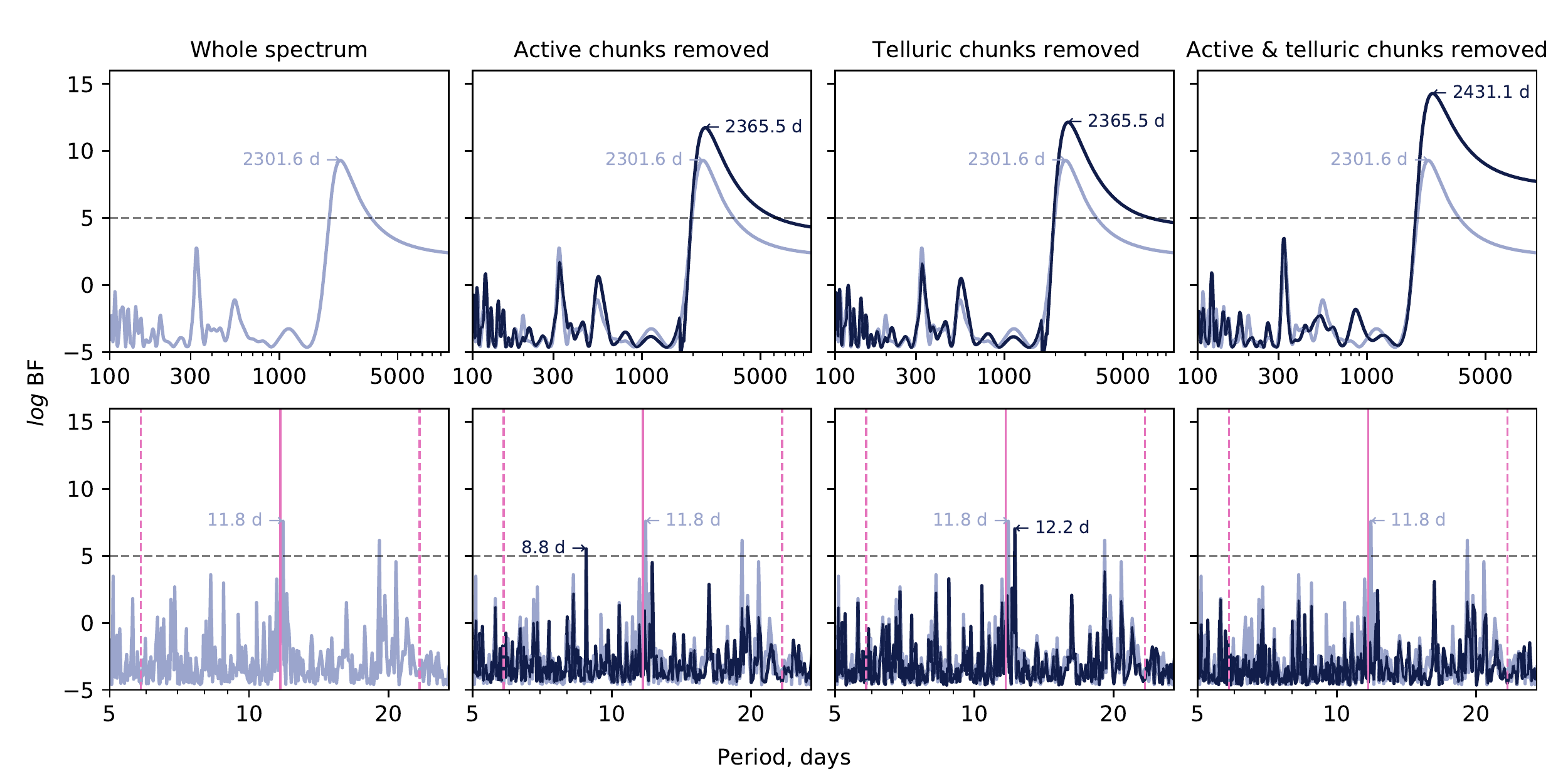}
        \label{fig:HD22049-bfp}
    \end{subfigure}

    \caption{
        Bayes Factor periodograms of HD 22049 radial velocity measurements (N=104).
        {\it From left to right:} original RV computed using the whole spectrum, RV with activity sensitive lines excluded, RV with telluric lines excluded and the RV with both active and telluric lines excluded.
        The {\it top} panels shows the periodograms centred at the planet's period, {\it bottom} panels shows the residuals after the planet is removed (centred at the stellar rotation period).
        The periodogram of the original data set is shown on all panels for comparison.
        Solid vertical line indicates the rotation period of the star and the double and half the period are indicated with vertical dashed lines.
        The periodograms in the top panel include the long period planet and a small 1 year peak.
    }
\label{fig:HD22049}
\end{figure*}

Figure \ref{fig:HD22049} shows BFPs of the computed data sets--from left to right: original radial velocity computed using the ``whole spectrum'' (light colour, left panels) , the one with activity sensitive lines excluded, the one with telluric lines excluded and the one with both active and telluric lines excluded from the RV computation (``partial spectra'').
The top panels shows a periodogram of the original data set (centred at the planet's period), while the bottom panels shows the residuals after the planet is removed (centred at the stellar rotation period).
The stellar rotation period of 11.68 days \citep[measured using Ca H\&K lines,][]{1996ApJ...466..384D} is marked with a solid vertical line.
As the harmonics of a rotation period might be falsely identified as a rotation period itself \citep{2020MNRAS.491.5216G}, we mark both double and one half periods with vertical dashed lines for reference.
The corresponding most significant signals are marked on the plots.
The periodogram of the original data set is shown on all panels for comparison.
In the latter case, the uncertainties on the individual radial velocity measurements estimated by the pipeline are slightly increased due to decreased amount of the spectrum used -- the median value is increased by about 10 per cent, from 1.02 \mps{} to 1.14 \mps.
This increase is insignificant compared to the amplitudes of the signals we are investigating.

The root-mean-square (RMS) of the data set is 12.5 \mps{} and it is slightly reduced to 12.1, 12.2 and 12.4 \mps{} by removing active, telluric and both sets of lines, respectively.

The first significant signal found is the planet.
The period is slightly shorter than the one found by \citet{2019AJ....157...33M}, but it is expected, as our data set only spans one orbital period.
It is more significant in the radial velocity dataset where active or telluric lines have been removed.
In the data set with both removed, it is much more significant and the period is increased by approximately 130 days.
Other peaks do not change significantly, which leads to a relative improvement of the planetary signal of 20, 23, and 39 per cent relative to the original data set for the three cases presented.
We do not see the 3 year activity period in the data \citep{2020A&A...636A..49C}, but it is possible that it is present as a change in the planet's period after the active lines are removed.

In addition, by removing active lines we expect to reduce significance of the stellar rotation signal, which is present in the data as a peak at 11.8 days.
It is effectively removed from the data after active lines are excluded from the radial velocity calculation, but persists when telluric lines are removed, albeit with a slightly longer period of 12.2 days.

A signal at 19.1 days of an unknown origin is also present in the data set and its significance is also reduced when active lines are removed, as well as when the telluric lines are removed (but to a smaller degree).

\subsubsection{Monte Carlo analysis}
\label{sec:results:hd22049:mc}

\begin{figure}
    \centering
    \includegraphics[width=0.45\textwidth]{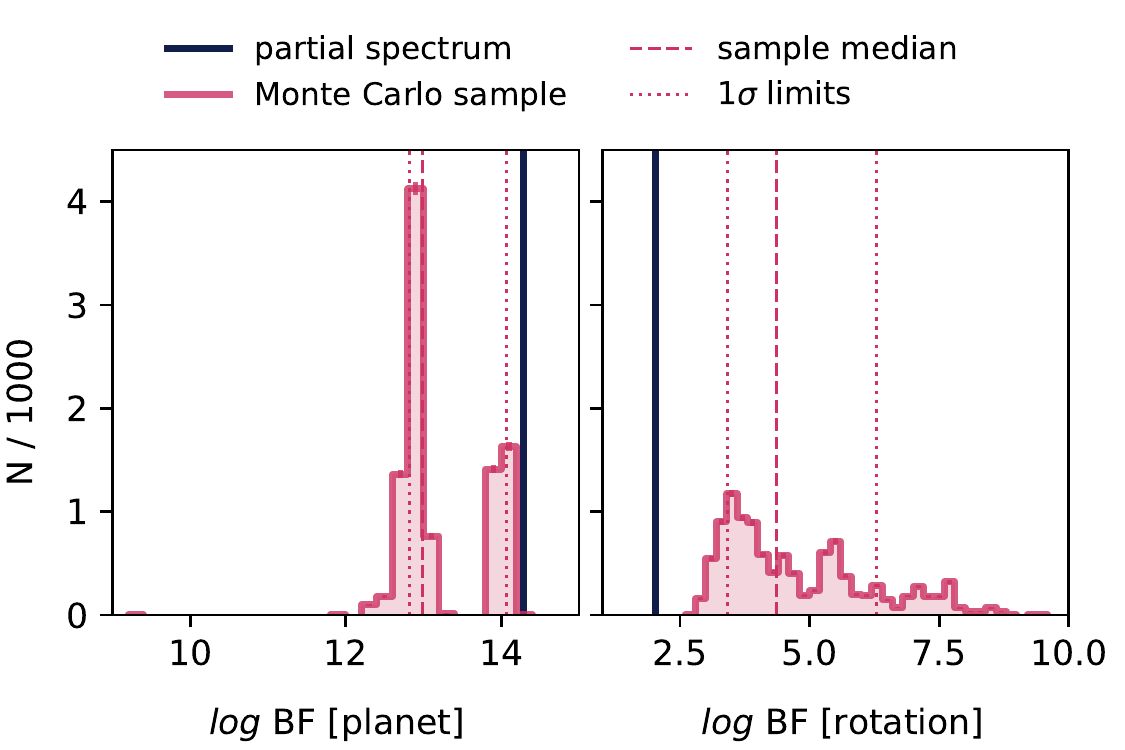}
    \caption{
        Monte Carlo simulation results for HD 22049.
        The {\it left} panel shows the distribution of the planetary signal peak height, {\it right} -- distribution of the stellar rotation signal peak height in the planet residuals.
        The median values and $1\sigma$ limits are plotted with vertical dashed and dotted lines, respectively.
        The value obtained from the data set with active and telluric lines removed is indicated with a solid vertical line.
        Poisson uncertainties are smaller than the thickness of the lines.
    }
    \label{fig:mc-hd22049}
\end{figure}

We perform a Monte Carlo test as described in Section \ref{sec:mc-intro}.
After running the pipeline 20,000 times, we obtained the distribution shown in Figure \ref{fig:mc-hd22049}.
The histograms show distributions of our the significance of the planetary signal ($ln$ BF) and significance of the stellar rotation signal after the planetary signal is removed.
Median values and 1$\sigma$ boundaries are indicated with red vertical lines.
The significances of the signals with the active and telluric lines removed are shown with solid vertical lines.
It can be seen that removing the lines from our list improves results.
That is, the planetary signal in the left-hand plot of Figure \ref{fig:mc-hd22049} has higher significance and the stellar rotation signature in the right-hand plot of Figure \ref{fig:mc-hd22049} is virtually gone.

The shapes of the distributions are highly non-Gaussian.
In case of the planetary signal the distribution has a duality, which might be caused by the fact that some of the chunks that are significantly affected are removed in one subsample but not the other.
The distribution also depends on the noise model used in the periodogram computation as the Bayes factor is estimated from the maximum likelihood ratio of the periodic model and the noise model \citep{Feng2016}.
In addition to removing wavelength dependent noise by removing the active chunks, a moving average noise model is adopted as optimal for removing time dependent noise.
When we performed a test trying to compare the partial spectra and the full spectrum, we found the baseline models between simulations are very different due to the different number of chunks and so a simple comparison is not possible. 
The moving average model performs differently between different sized data sets and generates quite different red noise properties so that that the simulations in Figure \ref{fig:mc-hd22049} can not simply be compared to the original full spectrum value.
We defer to Figure \ref{fig:HD22049} for direct comparison between the full and partial spectrum cases.
Figure \ref{fig:mc-hd22049} is a relative comparison between the Monte Carlo sample and our selection of lines to ensure that they perform better than random and provides a method for checking the robustness of a given Keplerian signal.

The rotational signal distribution also has at least two peaks, but produces a much more continuous distribution.
Our selection of lines, again, produces better results than the Monte Carlo sample, confirming the hypothesis of the selected lines being sensitive to activity (or, at least, rotation).
The value obtained with the list of lines lies outside the Monte Carlo distribution, which means that no random sample removed the same selection of lines.
This is not surprising, as the probability of randomly drawing specific 114 chunks out of 718 is $\approx 10^{-135}$.

The right panel of Figure \ref{fig:mc-hd22049} is also an example of usefulness of the Monte Carlo approach explored in this paper.
A Keplerian signal is expected to be uniform across the spectrum, producing a narrow histogram, because no matter which part of a spectrum is removed, the signal should be present at any wavelength.
The distribution of the rotational peak height that we see, on the other hand, is quite wide and about 62 per cent of the samples are below the detection limit, indicating considerable wavelength dependence.

The difference between the radial velocity values obtained from the whole spectrum and the ones with active and telluric lines excluded shows a small stellar rotation signal with an amplitude of 1.38 \mps.
In comparison, the 10 per cent increase in radial velocity uncertainties mentioned in Section \ref{sec:results:hd22049} is negligible.
The 365 days signal only changes slightly when the telluric lines are removed, because HIRES data already makes use of a telluric mask.
These lines are identified using a synthetic telluric atlas and corresponding pixels are masked out.
Evidently, these chunks still hold some contamination but not a strong periodic signal.

\subsection{HD 40979}
\label{sec:results:hd40979}



\begin{figure*}
    \centering
    \begin{subfigure}{.99\textwidth}
        \centering
        \includegraphics[width=\textwidth]{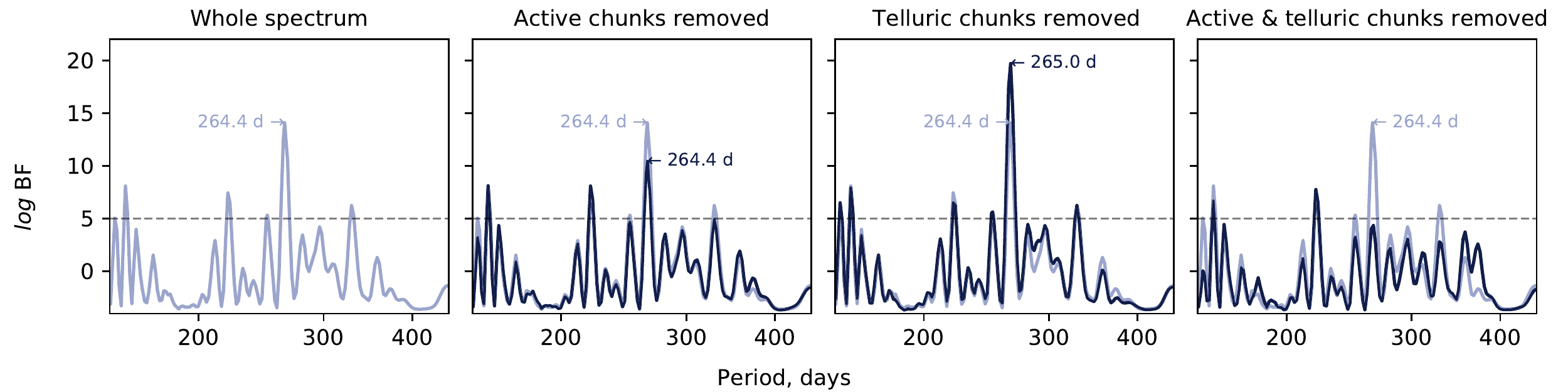}
        \label{fig:HD40979-bfp}
    \end{subfigure}

    \caption{
        Bayes Factor periodograms of HD 40979 radial velocity measurements (N=39).
        Same as Figure \ref{fig:HD22049}, but centred at the relevant periods.
    }
\label{fig:HD40979}
\end{figure*}

HD 40979 is a triple star system, the A component of which is an F8 star (see Figure \ref{fig:HD40979}), so more than a full spectral type away from $\alpha$ Cen B.
The B and C components are in a tight binary system separated from the main component by 262 arcsec \citep{2018A&A...616A...1G}--sufficiently far to not cause any light contamination.
In this case we investigate the effect of our analysis on an F star.
It is a moderately active star with log $R'_{HK}=-4.63$ and a rotation period of 7.8 days \citep[measured using Ca H\&K lines,][]{2017MNRAS.465.2734M}.
The system has a published planet with a period of $263.1 \pm 3.0$ days \citep{2003ApJ...586.1394F}.
Our sample has 39 observations, also taken with the HIRES instrument on Keck telescope, which provides good phase coverage for the planetary signal we are investigating.

This data set can be divided into two subsets divided by a CCD upgrade in 2004 \citep{2004SPIE.5496..178K, 2004SPIE.5492....1M}, but the difference between the two subsets did not make an impact on our analysis and we consider them jointly.
No significant rotation period is present in the radial velocity dataset or the subsets.

Figure \ref{fig:HD40979}, similar to Figure \ref{fig:HD22049}, shows periodograms of the original radial velocity computed using the whole spectrum, RV with activity sensitive lines excluded, RV with telluric lines excluded and the RV with both active and telluric lines excluded.
The planetary signal in the whole spectrum measurements is found to be $264.4 \pm 3.8$ days and thus the signal in the Keck dataset alone is consistent with various other published analyses \citep[e.g.][]{2003ApJ...586.1394F, 2006ApJ...646..505B} and with the combined data analysis of \citet{2009ApJS..182...97W} using 160 radial velocity data points from the Lick, Keck, HET and 2.7m telescopes.
When the active lines are removed, the signal becomes less significant, but still above detection threshold.
Removing telluric lines from the RV computation, on the other hand, improves the signal strength and the RMS around the planetary fit from 48.1 to 40.3 \mps.
Uniqueness of the signal in this case is increased by 33 per cent.
Evidently, in this case the active lines have less impact on the planetary signal compared to the telluric lines.
However, once both sets of lines are removed, the significance of the planetary signal is substantially reduced and it is no longer the highest peak.
This is analysed further in Section \ref{sec:results:impact} below.

Even though the 1 year period does not change significantly in the periodogram, it appears in the differential radial velocity.
The periodogram might not reflect its strength due to incomplete phase sampling and the fact that the planetary signal is almost an order of magnitude stronger.
The difference between the data sets with and without telluric lines included in the radial velocity calculation shows a 1 year signal with an amplitude of 14.68 \mps, which appears to be the origin of the improved RMS of the planetary fit on removal of telluric lines.
Although the reduction process filters telluric lines with iodine, telluric lines are not removed from the template as part of the reduction process.
Thus such a telluric signal is not entirely unexpected in a star with relatively weak spectral features.
The impact of tellurics is much more prominent compared to $\epsilon$ Eri, which is explained by the amount of Doppler information available in the stars' spectra and the relatively larger contribution of the telluric lines.

\subsubsection{Monte Carlo analysis}

\begin{figure}
    \centering
    \includegraphics[width=0.45\textwidth]{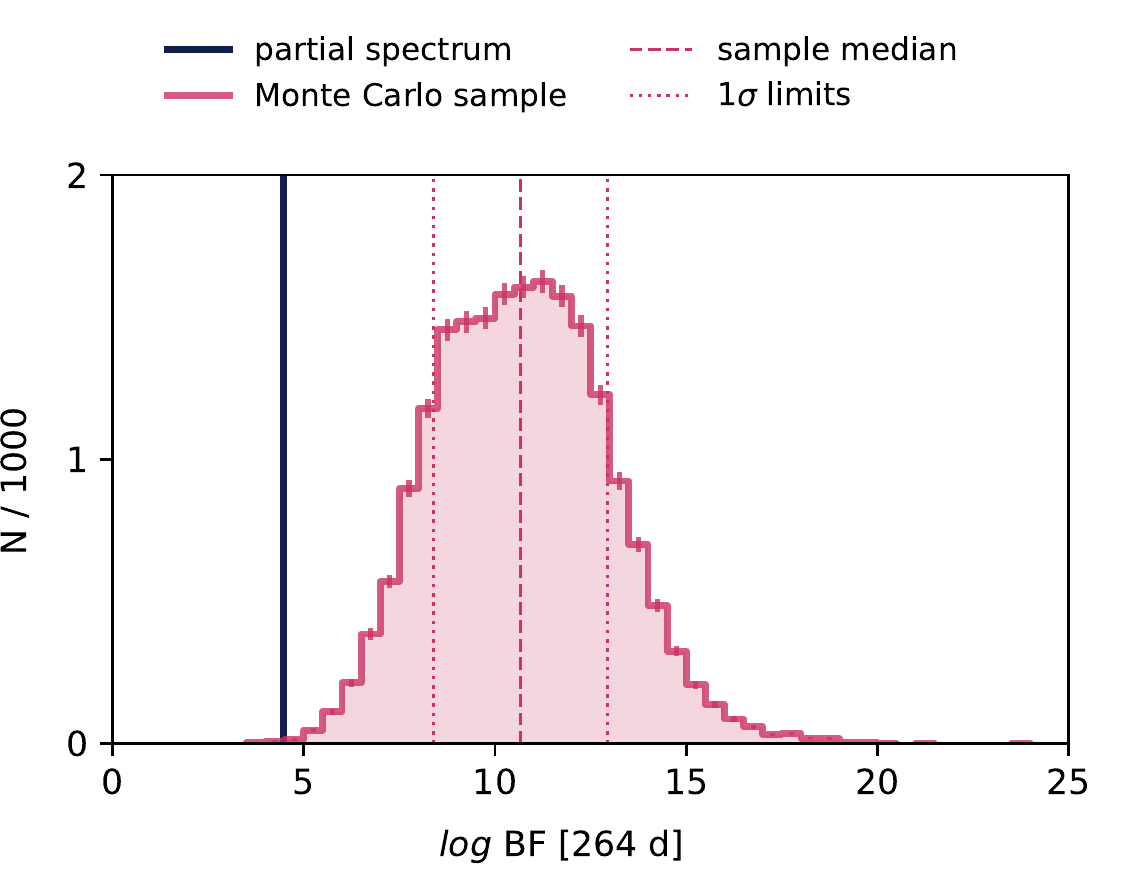}

    \caption{
        Monte Carlo simulation results for HD 40979.
        Same as Figure \ref{fig:mc-hd22049}, but the plot shows the distribution of the 265 days peak height.
        The error bars indicate Poisson uncertainties.
    }
    \label{fig:mc-hd40979}
\end{figure}

We perform the same test as in Section \ref{sec:results:hd22049:mc}.
The results of the test are presented in Figure \ref{fig:mc-hd40979}, which is similar to Figure \ref{fig:mc-hd22049}, but in this case a different periodic signal was investigated.
Again, the significance of the signal with the active and telluric lines removed is shown with a vertical line.
The figure shows the significance of the 264 days signal in the data set and the randomly sampled values fall between the two lines.
The distribution of the signal significance is rather wide, which is consistent with the wavelength dependence of the signal according to our hypothesis.
However, unlike the $\epsilon$ Eri rotation significance distribution (Figure \ref{fig:mc-hd22049}), 99.7 per cent of the samples are \textit{above} detectability threshold of $ln$ BF = 5.

In case of this star our line selection is not as successful as in case of a K dwarf and renders the planet undetected.
This means that stellar activity manifests itself in a very different set of spectral lines.

The distribution of the Monte Carlo simulation in Figure \ref{fig:mc-hd40979} might cast doubt on the existence of the planet, because Keplerian signals should be present in the whole spectrum.
Activity signals, on the other hand, are not evenly distributed across the spectrum, because they originate in individual spectral line variations.
We expect a signal that is present uniformly across the spectrum to produce a narrow distribution, which would indicate that no matter which part of the spectrum the radial velocity is measured from, the signal is present.
A signal with wavelength dependence will be detected or not depending on the wavelength selection, leading to a wide distribution of the peak height.
In this case, however, this hypothesis does not hold.

The signal could be attributed to a short activity cycle \citep{2019A&A...621A.136M}, but it is not evident in the $S$-index.
In addition, the activity-induced RV jitter is not expected to exceed 30 \mps{} based on its $R_{HK}$ and luminosity \citep{2020AJ....159..236L}, while the signal semi-amplitude is 109 \mps.
It is present in 99.7 per cent of the samples and the variation of the period across the samples is very low ($P = 264.69^{+0.25}_{-0.33}$ days, using 99.7 percentile uncertainties), which means that it is more likely to be of a planetary origin.

The signal becomes much less prominent after we remove the active lines and we hypothesize that they are quite strong features and are heavily relied upon for radial velocity information in hotter spectral types.
The wide distribution in Figure \ref{fig:mc-hd40979} can be explained by the much smaller amount of Doppler information in the spectrum of the star, as the metal lines in F stars are weaker than in K stars \citep[e.g.][]{2015RAA....15.1137L}, which is reduced even further after removing active and telluric chunks.
Removing a large part of the spectrum, especially some of the strongest lines, will introduce a lot of scatter and prevent signal detection.
However, there isn't a direct correlation between the amount of Doppler information in a spectrum and the planetary signal significance in this case.
This effect is discussed further and quantified in the following section.

\subsection{Impact on radial velocity precision}
\label{sec:results:impact}

As mentioned briefly in Section \ref{sec:actlines}, the active line selection is biased towards deep lines and removing them significantly reduces the amount of Doppler information extracted from a spectrum.
To quantify the impact on radial velocity uncertainties, we measure the amount of spectral gradient in each chunk of the stellar template used to measure the radial velocities.
This ``stellar'' template is a high signal-to-noise version of a star observed without iodine.
As we are interested in comparing the values, we measure it simply as a total absolute numerical derivative of the spectral range.
To reflect the chunk's contribution to the final radial velocity calculation, the values are then weighted using the weights provided by the pipeline \citep{1996pasp..108..500b}:

\begin{equation}
gradient_{chunk} = w_{chunk} \sum_{i=0}^{n} \frac{\lvert \Delta F_i \rvert}{\Delta \lambda_i},
\end{equation}

where $w_{chunk}$ is the weight of the chunk, $n$ is the number of pixels in a chunk, $\Delta F_i$ is the flux change in the $i$-th pixel, and $\Delta \lambda_i$ is the wavelength size of the $i$-th pixel.

\begin{table}
    \caption{
        Median and total weighted gradient in the stellar templates.
    }
    \label{tab:gradient-summary}
    \begin{tabular}{p{0.4\columnwidth}p{0.22\columnwidth}p{0.22\columnwidth}}
        \toprule
            Median chunk gradient: & HD 22049 & HD 40979 \\
        \midrule
            All chunks      & 0.683 & 0.051 \\
            Active chunks   & 1.300 & 0.082 \\
            Telluric chunks & 0.464 & 0.039 \\
        \bottomrule
    \end{tabular}
    \begin{tabular}{p{0.4\columnwidth}p{0.22\columnwidth}p{0.22\columnwidth}}

        \addlinespace[4ex]
            Total weighted gradient: & HD 22049 & HD 40979 \\
        \midrule
            All chunks               & 1005.4         & 119.6 \\
            Active chunks excluded   & 636.7 (63.3\%) & 85.9 (71.8\%) \\
            Telluric chunks excluded & 841.2 (83.7\%) & 94.4 (79.0\%) \\
            Both excluded            & 498.4 (49.6\%) & 62.2 (52.0\%) \\
        \bottomrule
    \end{tabular}
\end{table}

The results are summarised in Table \ref{tab:gradient-summary}.
The upper part of the table shows the median amount of weighted gradient per chunk in the whole spectrum as well as in the chunks containing active and telluric lines.
In both stars, the active lines we selected carry much more doppler information than the average across the spectrum.
Telluric chunks, on the other hand, are downweighted and therefore carry less effective Doppler information.
The bottom part of the table contains the total amount of weighted gradients in the full spectra and with the selected chunks removed.

This calculation highlights the difference between the two stars considered in this paper -- the sum of gradient across all chunks in HD 22049 (K2) template is almost an order of magnitude higher than in HD 40979 (F8).
Even with all active and telluric chunks removed from HD 22049 spectrum, it still contains four times more Doppler information than the full spectrum of HD 40979.
The relative reduction in weighted gradient is similar for both stars and so apparently does not bear out the suggestion that an F star losses proportionally more information.

\begin{figure*}
    \centering
    \begin{subfigure}{0.49\textwidth}
        \centering
        \includegraphics[width=\textwidth]{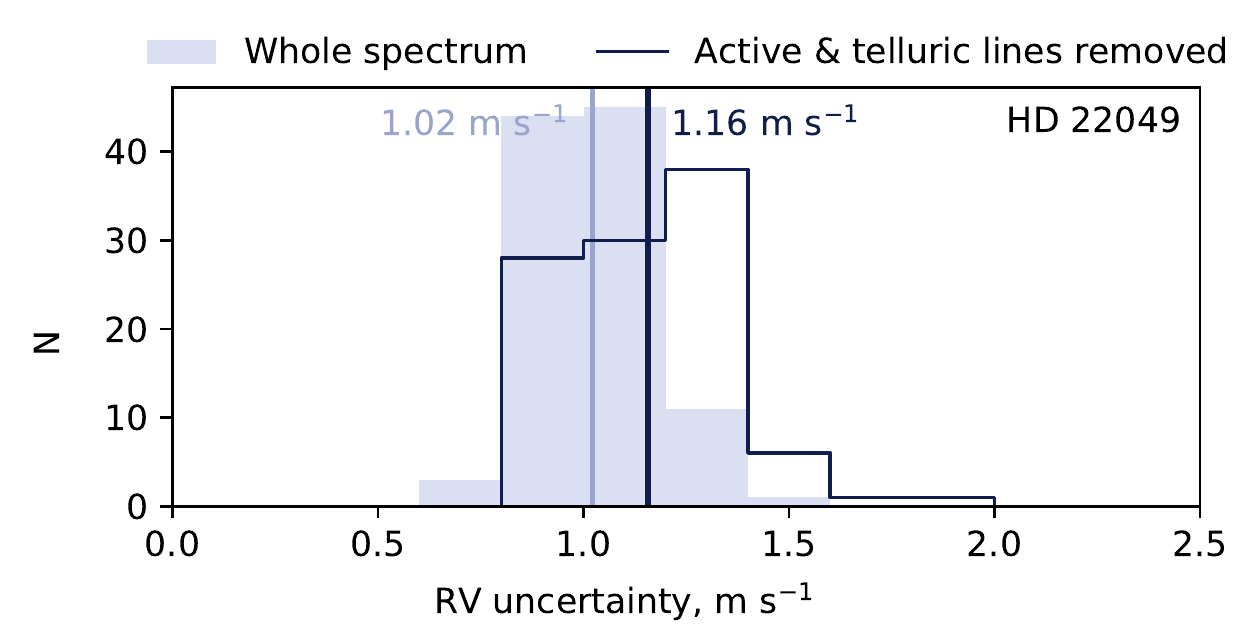}
        \label{fig:smth}
    \end{subfigure}
    \begin{subfigure}{0.49\textwidth}
        \centering
        \includegraphics[width=\textwidth]{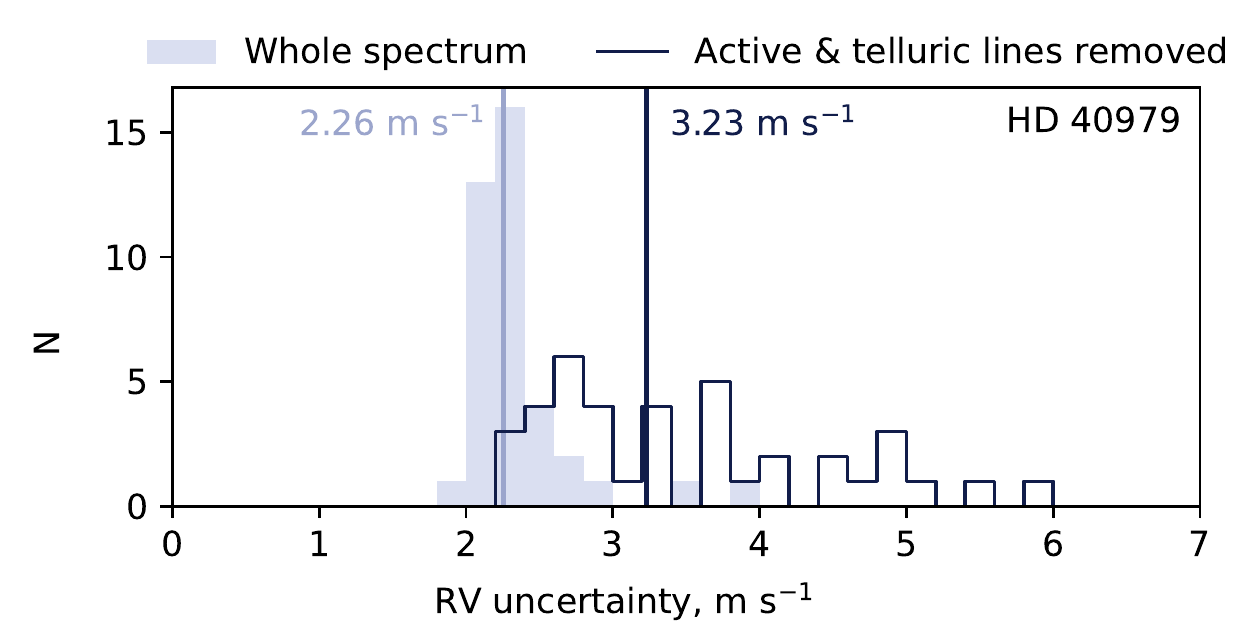}
        \label{fig:smth}
    \end{subfigure}
    \caption{
        Radial velocity uncertainty distribution for all observations of HD 22049 \textit{(left)} and HD 40979 \textit{(right)}.
        The histograms show the uncertainties estimated by the pipeline using all spectral chunks \textit{(solid bars)} and using only the chunks that do not contain the active and telluric lines in our lists \textit{(step line)}.
        The median values of each distribution are marked with vertical lines.
    }
\label{fig:uncertainties}
\end{figure*}

In terms of the radial velocity uncertainty estimated by the pipeline, however, it seems that the reduction of radial velocity information from removing wavelength chunks reaches a threshold where the uncertainties increase due to a loss of radial velocity information and the impact of removing ``active'' lines from the two different spectral types is illuminated.
Figure \ref{fig:uncertainties} shows the distribution of uncertainties in all radial velocity measurements for both stars.
Using the whole spectrum, both distributions are relatively narrow with median values of 1.02 \mps{} in case of HD 22049 and 2.26 \mps{} in case of HD 40979.
This difference becomes greater when the active and telluric chunks are removed from the calculation, as the F8 star is much more affected and its 
distribution of uncertainties becomes much broader unlike the K2 star.
This level of radial velocity uncertainty as well as the relative difference in velocity between spectral types K2 and F8 is also consistent with the expectations of studies investigating Doppler information content as a function of spectral type and range of radial velocities considered, e.g., table 2 of \citet{2020ApJS..247...11R}.

This test highlights that, while the method of removing active lines is appropriate for the stars with a wealth of deep spectral lines, one must be cautious about removing parts of the spectrum in stars with less Doppler information.
As we have shown above, the hypothesis of a relatively narrow distribution of wavelength dependence for a Doppler signal significance is not valid in case of greatly decreased radial velocity precision.

\subsection{Extending to other spectral types}
\label{sec:results:extending}

As we have shown in the previous section, the list of lines we used is sensitive to the spectral type.
Deriving it for a K star required a very specific data set (as described in Section \ref{sec:actlines}) that might not be currently available for many stars.

An alternative approach to identifying contaminating chunks in the iodine data would be the analysis of the Monte Carlo simulation results.
Suppose we have a star with a strong planetary signal and an iodine data set that contains it.
The star of choice needs to show at least a moderate level of activity variation as well.
We compute the Monte Carlo sample by removing random chunks of the spectrum and inspect the planetary signal significance in each of them (similar to Figures \ref{fig:mc-hd22049} and \ref{fig:mc-hd40979}).
The values will have a range of signal significance, in the worst case--from a significant detection to a non-detection.
If we compare which chunks were removed in the samples with the planet detected (Subsample A) to the ones with a non-detection (Subsample B), we will find the chunks that hinder the planet detection.
For instance, if in Subsample A a certain chunk was removed 100 per cent of the time and in Subsample B the same chunk was not removed at all, it means that it contains data that prevents the planet from being detected.
The resulting values of the spectrum will not be simply a list of active or inactive chunks, but measures of how contaminated the chunks are.
One could either use it to select top $N$ chunks to remove most of the contaminating noise in the spectra, or use the values as a weighting mask for the radial velocity computation.
Due to the random nature of this exercise and the number of combinations, the subsamples need to be quite big.

This approach does not require very special conditions for the data, only the presence of a well detected planet and some level of stellar activity variation.
The spectral weighting masks can be derived for all the spectral types with available data.

\section{Discussion}

We demonstrated how removing the parts of the spectrum that are more affected by stellar activity can remove the signals associated with it.
The approach improved the RMS, but we emphasise the fact that it is more important to remove the unwanted signals.
Removing the ``active'' chunks from the spectrum should work best for the most active stars, where they might introduce spurious signals.
For the less active stars, where these chunks are not dominated by activity, weighting might be a better solution, as it would reduce the impact of activity while preserving the radial velocity information and not increasing errors due to the removal of different spectral chunks.

In addition to increasing confidence in the existence of a planetary signal, one can compute a more reliable uncertainty estimate from the data and inform about ``problematic'' chunks of the spectrum affecting the radial velocity measurements, such as cosmic rays, telluric features etc.
For instance, computing multiple data sets from observations using various selections of spectral lines will give an estimate of the non-keplerian signals' contribution.

We select the active lines using period of elevated activity variation in $\alpha$ Cen B, and the list might not be representative.
If the rotational variation in activity indicators during that epoch was caused, for instance, by bright plages, the set of varying lines might differ from the one appearing in a more ``quiet'' state.
This needs to be investigated further using additional data and solar observations \citep[e.g.][]{2020arXiv200409830T}.

It is worth noting that the list is derived using a K1V star spectrum and is most applicable to close spectral types.
As different spectral types will have different spectral lines sensitive to activity-induced variation, additional study of a sample of stars would be useful to generalise the approach.
This would require carefully selecting data sets with high signal-to noise ratio, spectral resolution, activity variation and sampling.

These lines were shown to not be effective in case of HD 40979 in Section \ref{sec:results:hd40979}, due to a limited number of strong absorption features in the spectra.
Weighting might preserve more radial velocity information in the lines and produce better results for earlier spectral types.
Removing telluric lines from the radial velocity calculation, however, improved the RMS of the planetary signal substantially.

We highlight that Monte Carlo and bootstrapping methods are powerful tools and can be used effectively in exoplanet research.
In addition to increasing confidence in the existence of a planetary signal, one can compute a more reliable uncertainty estimate from the data and inform about ``problematic'' chunks of the spectrum affecting the radial velocity measurements, such as cosmic rays, telluric features etc.
For instance, computing multiple data sets from observations using various selections of spectral lines will give an estimate of the non-keplerian signals' contribution.
Activity signals are not robust against random spectral ranges removal, as the activity sensitive spectral lines are not uniformly distributed across the spectrum.
These lines also vary to different degrees and show a number of different shape variations, which lead to different radial velocity effects.
Keplerian signals, on the other hand, come from a star moving as a whole and affect the whole spectrum in the same way.
When one measures the significance of an exoplanet detection from a random selection of spectral chunks, the shape of the resulting distribution will be wide for signals that are not uniformly present in the whole spectrum and narrow for the ones that are.
Further work will be required to present this as a simple metric but when a signal presents as relatively low significance along with a wide distribution of wavelength dependence this is an indication that the signal is not detected robustly.

Additional tests must be conducted to investigate the distributions of the signal significance, as the problem is not a simple one.
The results are dependent on significance of the signal and might be sensitive to line removal in general.
For instance, $\epsilon$ Eri b signal is not well constrained due to the data only covering one period of the planet.
In this work we have only explored the wavelength dimension of stellar activity data, however, radial velocity data is also time correlated.
Further investigations are required to understand and mitigate the impact of active lines and how the removal of stellar lines simultaneously changes the time correlation of the data.

Iodine data is particularly easy to use for these methods in post-processing, as it allows retrieval of hundreds of radial velocity measurements from a single spectrum.
Other radial velocity pipelines provide a smaller number of spectral orders by default \citep[e.g. 72 in case of HARPS,][]{2000SPIE.4008..582P}, which means most radial velocity packages for HARPS-like data blend many lines into one radial velocity measurement per order.
However, specific lines can also be removed simply by rejecting pixels containing that line in the mask used to build a CCF.
Whereas in this work we are removing the full chunk which represents 2\AA{}, the CCF method is much more flexible and can remove parts of a spectrum as small as 0.2\AA{} (at 5000\AA{} for a HARPS spectra).
For instance, a 1 year signal was corrected in HARPS data using this approach in \citet{2015ApJ...808..171D}, which shows that the methods explored in this work are widely applicable to any data and would be more effective if applied in pre-processing.

\section*{Acknowledgements}

We would like to thank Joanna Ramasawmy for her expertise and fruitful discussions that made this paper possible.
We also thank the anonymous referee for their insights which led to substantial improvements in this manuscript.

ML acknowledges financial support from University of Hertfordshire PhD studentship.
HJ acknowledges support from the UK Science and Technology Facilities Council [ST/R006598/1].

This research made extensive use of \textsc{Agatha} software \citep{feng17agatha}, 
\textsc{numpy} \citep{van2011numpy}, 
\textsc{scipy} \citep{jones_scipy_2001}, 
and \textsc{matplotlib}, a Python library for publication quality graphics \citep{Hunter:2007}.

\section*{Data availability statement}

The data underlying this article are available on \textsc{Github}, at \url{https://github.com/timberhill/iodine-monte-carlo}.
The radial velocity data sets were computed using iodine data pipeline \citep{1996pasp..108..500b}.
The periodograms were computed using \textsc{Agatha} software \citep{feng17agatha} available on \textsc{Github} at \url{https://github.com/phillippro/agatha} and \url{https://github.com/phillippro/Agatha2.0}.


\bibliographystyle{mnras}
\bibliography{paper}

\begin{thebibliography}{}
\makeatletter
\relax
\def\mn@urlcharsother{\let\do\@makeother \do\$\do\&\do\#\do\^\do\_\do\%\do\~}
\def\mn@doi{\begingroup\mn@urlcharsother \@ifnextchar [ {\mn@doi@}
  {\mn@doi@[]}}
\def\mn@doi@[#1]#2{\def\@tempa{#1}\ifx\@tempa\@empty \href
  {http://dx.doi.org/#2} {doi:#2}\else \href {http://dx.doi.org/#2} {#1}\fi
  \endgroup}
\def\mn@eprint#1#2{\mn@eprint@#1:#2::\@nil}
\def\mn@eprint@arXiv#1{\href {http://arxiv.org/abs/#1} {{\tt arXiv:#1}}}
\def\mn@eprint@dblp#1{\href {http://dblp.uni-trier.de/rec/bibtex/#1.xml}
  {dblp:#1}}
\def\mn@eprint@#1:#2:#3:#4\@nil{\def\@tempa {#1}\def\@tempb {#2}\def\@tempc
  {#3}\ifx \@tempc \@empty \let \@tempc \@tempb \let \@tempb \@tempa \fi \ifx
  \@tempb \@empty \def\@tempb {arXiv}\fi \@ifundefined
  {mn@eprint@\@tempb}{\@tempb:\@tempc}{\expandafter \expandafter \csname
  mn@eprint@\@tempb\endcsname \expandafter{\@tempc}}}

\bibitem[\protect\citeauthoryear{Baluev}{Baluev}{2013}]{BALUEV201318}
Baluev R.~V.,  2013, \mn@doi [Astronomy and Computing]
  {https://doi.org/10.1016/j.ascom.2013.07.001}, 2, 18

\bibitem[\protect\citeauthoryear{{Barnes}}{{Barnes}}{2007}]{2007ApJ...669.1167B}
{Barnes} S.~A.,  2007, \mn@doi [\apj] {10.1086/519295}, \href
  {https://ui.adsabs.harvard.edu/abs/2007ApJ...669.1167B} {669, 1167}

\bibitem[\protect\citeauthoryear{{Butler}, {Marcy}, {Williams}, {McCarthy},
  {Dosanjh}  \& {Vogt}}{{Butler} et~al.}{1996}]{1996pasp..108..500b}
{Butler} R.~P.,  {Marcy} G.~W.,  {Williams} E.,  {McCarthy} C.,  {Dosanjh} P.,
   {Vogt} S.~S.,  1996, \mn@doi [\pasp] {10.1086/133755}, \href
  {https://ui.adsabs.harvard.edu/abs/1996PASP..108..500B} {108, 500}

\bibitem[\protect\citeauthoryear{{Butler}, {Marcy}, {Fischer}, {Brown},
  {Contos}, {Korzennik}, {Nisenson}  \& {Noyes}}{{Butler}
  et~al.}{1999}]{1999ApJ...526..916B}
{Butler} R.~P.,  {Marcy} G.~W.,  {Fischer} D.~A.,  {Brown} T.~M.,  {Contos}
  A.~R.,  {Korzennik} S.~G.,  {Nisenson} P.,   {Noyes} R.~W.,  1999, \mn@doi
  [\apj] {10.1086/308035}, \href
  {https://ui.adsabs.harvard.edu/abs/1999ApJ...526..916B} {526, 916}

\bibitem[\protect\citeauthoryear{{Butler}, {Vogt}, {Marcy}, {Fischer},
  {Wright}, {Henry}, {Laughlin}  \& {Lissauer}}{{Butler}
  et~al.}{2004}]{2004ApJ...617..580B}
{Butler} R.~P.,  {Vogt} S.~S.,  {Marcy} G.~W.,  {Fischer} D.~A.,  {Wright}
  J.~T.,  {Henry} G.~W.,  {Laughlin} G.,   {Lissauer} J.~J.,  2004, \mn@doi
  [\apj] {10.1086/425173}, \href
  {https://ui.adsabs.harvard.edu/abs/2004ApJ...617..580B} {617, 580}

\bibitem[\protect\citeauthoryear{{Butler} et~al.,}{{Butler}
  et~al.}{2006}]{2006ApJ...646..505B}
{Butler} R.~P.,  et~al., 2006, \mn@doi [\apj] {10.1086/504701}, \href
  {https://ui.adsabs.harvard.edu/abs/2006ApJ...646..505B} {646, 505}

\bibitem[\protect\citeauthoryear{{Butler} et~al.,}{{Butler}
  et~al.}{2017}]{2017AJ....153..208B}
{Butler} R.~P.,  et~al., 2017, \mn@doi [\aj] {10.3847/1538-3881/aa66ca}, \href
  {https://ui.adsabs.harvard.edu/abs/2017AJ....153..208B} {153, 208}

\bibitem[\protect\citeauthoryear{{Carleo} et~al.,}{{Carleo}
  et~al.}{2020}]{2020A&A...638A...5C}
{Carleo} I.,  et~al., 2020, \mn@doi [\aap] {10.1051/0004-6361/201937369}, \href
  {https://ui.adsabs.harvard.edu/abs/2020A&A...638A...5C} {638, A5}

\bibitem[\protect\citeauthoryear{{Claudi} et~al.,}{{Claudi}
  et~al.}{2017}]{2017EPJP..132..364C}
{Claudi} R.,  et~al., 2017, \mn@doi [European Physical Journal Plus]
  {10.1140/epjp/i2017-11647-9}, \href
  {https://ui.adsabs.harvard.edu/abs/2017EPJP..132..364C} {132, 364}

\bibitem[\protect\citeauthoryear{{Coffaro} et~al.,}{{Coffaro}
  et~al.}{2020}]{2020A&A...636A..49C}
{Coffaro} M.,  et~al., 2020, \mn@doi [\aap] {10.1051/0004-6361/201936479},
  \href {https://ui.adsabs.harvard.edu/abs/2020A&A...636A..49C} {636, A49}

\bibitem[\protect\citeauthoryear{{Cretignier}, {Dumusque}, {Allart}, {Pepe}  \&
  {Lovis}}{{Cretignier} et~al.}{2020}]{2020A&A...633A..76C}
{Cretignier} M.,  {Dumusque} X.,  {Allart} R.,  {Pepe} F.,   {Lovis} C.,  2020,
  \mn@doi [\aap] {10.1051/0004-6361/201936548}, \href
  {https://ui.adsabs.harvard.edu/abs/2020A&A...633A..76C} {633, A76}

\bibitem[\protect\citeauthoryear{{Cunha}, {Santos}, {Figueira}, {Santerne},
  {Bertaux}  \& {Lovis}}{{Cunha} et~al.}{2014}]{2014A&A...568A..35C}
{Cunha} D.,  {Santos} N.~C.,  {Figueira} P.,  {Santerne} A.,  {Bertaux} J.~L.,
   {Lovis} C.,  2014, \mn@doi [\aap] {10.1051/0004-6361/201423723}, \href
  {https://ui.adsabs.harvard.edu/abs/2014A&A...568A..35C} {568, A35}

\bibitem[\protect\citeauthoryear{{Donahue}, {Saar}  \& {Baliunas}}{{Donahue}
  et~al.}{1996}]{1996ApJ...466..384D}
{Donahue} R.~A.,  {Saar} S.~H.,   {Baliunas} S.~L.,  1996, \mn@doi [\apj]
  {10.1086/177517}, \href
  {https://ui.adsabs.harvard.edu/abs/1996ApJ...466..384D} {466, 384}

\bibitem[\protect\citeauthoryear{{Dumusque}}{{Dumusque}}{2018}]{2018A&A...620A..47D}
{Dumusque} X.,  2018, \mn@doi [\aap] {10.1051/0004-6361/201833795}, \href
  {https://ui.adsabs.harvard.edu/abs/2018A&A...620A..47D} {620, A47}

\bibitem[\protect\citeauthoryear{{Dumusque} et~al.,}{{Dumusque}
  et~al.}{2012}]{2012Natur.491..207D}
{Dumusque} X.,  et~al., 2012, \mn@doi [\nat] {10.1038/nature11572}, \href
  {https://ui.adsabs.harvard.edu/abs/2012Natur.491..207D} {491, 207}

\bibitem[\protect\citeauthoryear{{Dumusque}, {Pepe}, {Lovis}  \&
  {Latham}}{{Dumusque} et~al.}{2015}]{2015ApJ...808..171D}
{Dumusque} X.,  {Pepe} F.,  {Lovis} C.,   {Latham} D.~W.,  2015, \mn@doi [\apj]
  {10.1088/0004-637X/808/2/171}, \href
  {https://ui.adsabs.harvard.edu/abs/2015ApJ...808..171D} {808, 171}

\bibitem[\protect\citeauthoryear{{Duncan} et~al.,}{{Duncan}
  et~al.}{1991}]{1991ApJS...76..383D}
{Duncan} D.~K.,  et~al., 1991, \mn@doi [\apjs] {10.1086/191572}, \href
  {https://ui.adsabs.harvard.edu/abs/1991ApJS...76..383D} {76, 383}

\bibitem[\protect\citeauthoryear{Feng, Tuomi, Jones, Butler  \& Vogt}{Feng
  et~al.}{2016}]{Feng2016}
Feng F.,  Tuomi M.,  Jones H. R.~A.,  Butler R.~P.,   Vogt S.,  2016, \mn@doi
  [Monthly Notices of the Royal Astronomical Society] {10.1093/mnras/stw1478},
  461, 2440

\bibitem[\protect\citeauthoryear{{Feng}, {Tuomi}, {Jones}, {Barnes},
  {Anglada-Escud{\'e}}, {Vogt}  \& {Butler}}{{Feng}
  et~al.}{2017a}]{2017AJ....154..135F}
{Feng} F.,  {Tuomi} M.,  {Jones} H.~R.~A.,  {Barnes} J.,  {Anglada-Escud{\'e}}
  G.,  {Vogt} S.~S.,   {Butler} R.~P.,  2017a, \mn@doi [\aj]
  {10.3847/1538-3881/aa83b4}, \href
  {https://ui.adsabs.harvard.edu/abs/2017AJ....154..135F} {154, 135}

\bibitem[\protect\citeauthoryear{{Feng}, {Tuomi}  \& {Jones}}{{Feng}
  et~al.}{2017b}]{feng17agatha}
{Feng} F.,  {Tuomi} M.,   {Jones} H.~R.~A.,  2017b, \mn@doi [\mnras]
  {10.1093/mnras/stx1126}, \href
  {http://adsabs.harvard.edu/abs/2017MNRAS.470.4794F} {470, 4794}

\bibitem[\protect\citeauthoryear{{Feng}, {Tuomi}  \& {Jones}}{{Feng}
  et~al.}{2017c}]{2017A&A...605A.103F}
{Feng} F.,  {Tuomi} M.,   {Jones} H.~R.~A.,  2017c, \mn@doi [\aap]
  {10.1051/0004-6361/201730406}, \href
  {https://ui.adsabs.harvard.edu/abs/2017A&A...605A.103F} {605, A103}

\bibitem[\protect\citeauthoryear{{Fischer} et~al.,}{{Fischer}
  et~al.}{2003}]{2003ApJ...586.1394F}
{Fischer} D.~A.,  et~al., 2003, \mn@doi [\apj] {10.1086/367889}, \href
  {https://ui.adsabs.harvard.edu/abs/2003ApJ...586.1394F} {586, 1394}

\bibitem[\protect\citeauthoryear{{Gaia Collaboration} et~al.,}{{Gaia
  Collaboration} et~al.}{2018}]{2018A&A...616A...1G}
{Gaia Collaboration} et~al., 2018, \mn@doi [\aap]
  {10.1051/0004-6361/201833051}, \href
  {https://ui.adsabs.harvard.edu/abs/2018A&A...616A...1G} {616, A1}

\bibitem[\protect\citeauthoryear{{Giacobbe} et~al.,}{{Giacobbe}
  et~al.}{2020}]{2020MNRAS.491.5216G}
{Giacobbe} P.,  et~al., 2020, \mn@doi [\mnras] {10.1093/mnras/stz3364}, \href
  {https://ui.adsabs.harvard.edu/abs/2020MNRAS.491.5216G} {491, 5216}

\bibitem[\protect\citeauthoryear{Gray}{Gray}{2005}]{Gray_2005}
Gray D.~F.,  2005, \mn@doi [Publications of the Astronomical Society of the
  Pacific] {10.1086/430412}, 117, 711

\bibitem[\protect\citeauthoryear{Hunter}{Hunter}{2007}]{Hunter:2007}
Hunter J.~D.,  2007, Computing In Science \& Engineering, 9, 90

\bibitem[\protect\citeauthoryear{Jones, Oliphant, Peterson  \& Others}{Jones
  et~al.}{2001}]{jones_scipy_2001}
Jones E.,  Oliphant T.,  Peterson P.,   Others 2001, {SciPy}: Open source
  scientific tools for Python, \url {http://www.scipy.org/}

\bibitem[\protect\citeauthoryear{{Keenan} \& {McNeil}}{{Keenan} \&
  {McNeil}}{1989}]{1989ApJS...71..245K}
{Keenan} P.~C.,  {McNeil} R.~C.,  1989, \mn@doi [\apjs] {10.1086/191373}, \href
  {https://ui.adsabs.harvard.edu/abs/1989ApJS...71..245K} {71, 245}

\bibitem[\protect\citeauthoryear{{Kibrick}, {Wright}, {Allen}  \&
  {Clarke}}{{Kibrick} et~al.}{2004}]{2004SPIE.5496..178K}
{Kibrick} R.~I.,  {Wright} C.,  {Allen} S.~L.,   {Clarke} D.~A.,  2004, in
  {Lewis} H.,  {Raffi} G.,  eds,  Society of Photo-Optical Instrumentation
  Engineers (SPIE) Conference Series Vol. 5496, \procspie. pp 178--189,
  \mn@doi{10.1117/12.552404}

\bibitem[\protect\citeauthoryear{{Lisogorskyi}, {Jones}  \&
  {Feng}}{{Lisogorskyi} et~al.}{2019}]{2019MNRAS.485.4804L}
{Lisogorskyi} M.,  {Jones} H.~R.~A.,   {Feng} F.,  2019, \mn@doi [\mnras]
  {10.1093/mnras/stz694}, \href
  {https://ui.adsabs.harvard.edu/abs/2019MNRAS.485.4804L} {485, 4804}

\bibitem[\protect\citeauthoryear{{Liu} et~al.,}{{Liu}
  et~al.}{2015}]{2015RAA....15.1137L}
{Liu} C.,  et~al., 2015, \mn@doi [Research in Astronomy and Astrophysics]
  {10.1088/1674-4527/15/8/004}, \href
  {https://ui.adsabs.harvard.edu/abs/2015RAA....15.1137L} {15, 1137}

\bibitem[\protect\citeauthoryear{{Luhn}, {Wright}  \& {Isaacson}}{{Luhn}
  et~al.}{2020}]{2020AJ....159..236L}
{Luhn} J.~K.,  {Wright} J.~T.,   {Isaacson} H.,  2020, \mn@doi [\aj]
  {10.3847/1538-3881/ab775c}, \href
  {https://ui.adsabs.harvard.edu/abs/2020AJ....159..236L} {159, 236}

\bibitem[\protect\citeauthoryear{{Marcy} \& {Butler}}{{Marcy} \&
  {Butler}}{1992}]{1992PASP..104..270M}
{Marcy} G.~W.,  {Butler} R.~P.,  1992, \mn@doi [\pasp] {10.1086/132989}, \href
  {https://ui.adsabs.harvard.edu/abs/1992PASP..104..270M} {104, 270}

\bibitem[\protect\citeauthoryear{{Mawet} et~al.,}{{Mawet}
  et~al.}{2019}]{2019AJ....157...33M}
{Mawet} D.,  et~al., 2019, \mn@doi [\aj] {10.3847/1538-3881/aaef8a}, \href
  {https://ui.adsabs.harvard.edu/abs/2019AJ....157...33M} {157, 33}

\bibitem[\protect\citeauthoryear{{McLean} \& {Adkins}}{{McLean} \&
  {Adkins}}{2004}]{2004SPIE.5492....1M}
{McLean} I.~S.,  {Adkins} S.,  2004, in {Moorwood} A. F.~M.,  {Iye} M.,  eds,
  Society of Photo-Optical Instrumentation Engineers (SPIE) Conference Series
  Vol. 5492, \procspie. pp 1--12, \mn@doi{10.1117/12.552154}

\bibitem[\protect\citeauthoryear{{Mengel} et~al.,}{{Mengel}
  et~al.}{2017}]{2017MNRAS.465.2734M}
{Mengel} M.~W.,  et~al., 2017, \mn@doi [\mnras] {10.1093/mnras/stw2949}, \href
  {https://ui.adsabs.harvard.edu/abs/2017MNRAS.465.2734M} {465, 2734}

\bibitem[\protect\citeauthoryear{{Metcalfe}, {Henry}, {Kn{\"o}lker}  \&
  {Soderblom}}{{Metcalfe} et~al.}{2006}]{2006ESASP.624E.111M}
{Metcalfe} T.~S.,  {Henry} T.~J.,  {Kn{\"o}lker} M.,   {Soderblom} D.~R.,
  2006, in Proceedings of SOHO 18/GONG 2006/HELAS I, Beyond the spherical Sun.
  p.~111 (\mn@eprint {arXiv} {astro-ph/0609051})

\bibitem[\protect\citeauthoryear{{Mittag}, {Schmitt}, {Hempelmann}  \&
  {Schr{\"o}der}}{{Mittag} et~al.}{2019}]{2019A&A...621A.136M}
{Mittag} M.,  {Schmitt} J.~H.~M.~M.,  {Hempelmann} A.,   {Schr{\"o}der} K.~P.,
  2019, \mn@doi [\aap] {10.1051/0004-6361/201834319}, \href
  {https://ui.adsabs.harvard.edu/abs/2019A&A...621A.136M} {621, A136}

\bibitem[\protect\citeauthoryear{{Pepe} et~al.,}{{Pepe}
  et~al.}{2000}]{2000SPIE.4008..582P}
{Pepe} F.,  et~al., 2000, {HARPS: a new high-resolution spectrograph for the
  search of extrasolar planets}.
\procspie, pp 582--592, \mn@doi{10.1117/12.395516}

\bibitem[\protect\citeauthoryear{{Reiners} \& {Zechmeister}}{{Reiners} \&
  {Zechmeister}}{2020}]{2020ApJS..247...11R}
{Reiners} A.,  {Zechmeister} M.,  2020, \mn@doi [\apjs]
  {10.3847/1538-4365/ab609f}, \href
  {https://ui.adsabs.harvard.edu/abs/2020ApJS..247...11R} {247, 11}

\bibitem[\protect\citeauthoryear{{Thompson}, {Watson}, {de Mooij}  \&
  {Jess}}{{Thompson} et~al.}{2017}]{2017MNRAS.468L..16T}
{Thompson} A.~P.~G.,  {Watson} C.~A.,  {de Mooij} E.~J.~W.,   {Jess} D.~B.,
  2017, \mn@doi [\mnras] {10.1093/mnrasl/slx018}, \href
  {https://ui.adsabs.harvard.edu/abs/2017MNRAS.468L..16T} {468, L16}

\bibitem[\protect\citeauthoryear{{Thompson} et~al.,}{{Thompson}
  et~al.}{2020}]{2020arXiv200409830T}
{Thompson} A.~P.~G.,  et~al., 2020, arXiv e-prints, \href
  {https://ui.adsabs.harvard.edu/abs/2020arXiv200409830T} {p. arXiv:2004.09830}

\bibitem[\protect\citeauthoryear{{Tuomi}, {Jones}, {Barnes},
  {Anglada-Escud{\'e}}, {Butler}, {Kiraga}  \& {Vogt}}{{Tuomi}
  et~al.}{2018}]{2018AJ....155..192T}
{Tuomi} M.,  {Jones} H. R.~A.,  {Barnes} J.~R.,  {Anglada-Escud{\'e}} G.,
  {Butler} R.~P.,  {Kiraga} M.,   {Vogt} S.~S.,  2018, \mn@doi [\aj]
  {10.3847/1538-3881/aab09c}, \href
  {https://ui.adsabs.harvard.edu/abs/2018AJ....155..192T} {155, 192}

\bibitem[\protect\citeauthoryear{Van Der~Walt, Colbert  \& Varoquaux}{Van
  Der~Walt et~al.}{2011}]{van2011numpy}
Van Der~Walt S.,  Colbert S.~C.,   Varoquaux G.,  2011, Computing in Science \&
  Engineering, 13, 22

\bibitem[\protect\citeauthoryear{{Wise}, {Dodson-Robinson}, {Bevenour}  \&
  {Provini}}{{Wise} et~al.}{2018}]{2018AJ....156..180W}
{Wise} A.~W.,  {Dodson-Robinson} S.~E.,  {Bevenour} K.,   {Provini} A.,  2018,
  \mn@doi [\aj] {10.3847/1538-3881/aadd94}, \href
  {https://ui.adsabs.harvard.edu/abs/2018AJ....156..180W} {156, 180}

\bibitem[\protect\citeauthoryear{{Wittenmyer}, {Endl}, {Cochran}, {Levison}  \&
  {Henry}}{{Wittenmyer} et~al.}{2009}]{2009ApJS..182...97W}
{Wittenmyer} R.~A.,  {Endl} M.,  {Cochran} W.~D.,  {Levison} H.~F.,   {Henry}
  G.~W.,  2009, \mn@doi [\apjs] {10.1088/0067-0049/182/1/97}, \href
  {https://ui.adsabs.harvard.edu/abs/2009ApJS..182...97W} {182, 97}

\bibitem[\protect\citeauthoryear{{Xuesong Wang}, {Wright}, {MacQueen},
  {Cochran}, {Doss}, {Gibson}  \& {Schmitt}}{{Xuesong Wang}
  et~al.}{2020}]{2020PASP..132a4503X}
{Xuesong Wang} S.,  {Wright} J.~T.,  {MacQueen} P.,  {Cochran} W.~D.,  {Doss}
  D.~R.,  {Gibson} C.~A.,   {Schmitt} J.~R.,  2020, \mn@doi [\pasp]
  {10.1088/1538-3873/ab5021}, \href
  {https://ui.adsabs.harvard.edu/abs/2020PASP..132a4503X} {132, 014503}

\makeatother
\end{thebibliography}


\appendix
\section{Telluric lines list}
\label{sec:appendix}
\begin{table*}
    \begin{threeparttable}
        \centering
      \caption{
        List of telluric lines used in this work, compiled as described in Section \ref{sec:tellines},
        including their wavelengths and depths relative to a normalised continuum of 1.
    }
\begin{tabular}{p{1cm}p{1.5cm}|p{1cm}p{1.5cm}|p{1cm}p{1.5cm}|p{1cm}p{1.5cm}|p{1cm}p{1.5cm}}
        \toprule
      $\lambda$, \AA & Depth, \% &       $\lambda$, \AA & Depth, \% &       $\lambda$, \AA & Depth, \% &       $\lambda$, \AA & Depth, \% &       $\lambda$, \AA & Depth, \% \\
        \midrule
5030.14 & 2.12 & 5457.11 & 2.60 & 5883.60 & 4.45 & 5908.07 & 5.52 & 5941.22 & 8.60  \\
5035.19 & 2.12 & 5463.17 & 2.49 & 5883.78 & 3.05 & 5908.17 & 2.57 & 5941.67 & 3.84  \\
5035.21 & 2.32 & 5682.53 & 2.16 & 5883.82 & 3.08 & 5908.27 & 2.67 & 5941.88 & 23.47  \\
5035.36 & 2.37 & 5684.57 & 2.29 & 5883.87 & 2.28 & 5908.63 & 6.40 & 5942.00 & 2.87  \\
5035.94 & 2.16 & 5684.63 & 3.49 & 5884.69 & 13.18 & 5909.00 & 6.31 & 5942.05 & 2.99  \\
5036.09 & 2.16 & 5686.19 & 3.12 & 5885.00 & 3.35 & 5909.79 & 13.05 & 5942.43 & 17.72  \\
5036.40 & 2.99 & 5688.24 & 6.68 & 5886.44 & 3.25 & 5910.25 & 2.33 & 5943.07 & 2.12  \\
5036.44 & 2.39 & 5690.35 & 3.61 & 5886.77 & 19.79 & 5910.99 & 5.76 & 5943.10 & 2.06  \\
5037.99 & 5.34 & 5691.20 & 4.74 & 5887.13 & 7.36 & 5911.08 & 2.33 & 5943.22 & 11.39  \\
5038.42 & 5.16 & 5693.19 & 6.07 & 5887.47 & 4.02 & 5911.12 & 2.07 & 5943.37 & 22.36  \\
5038.96 & 2.38 & 5696.72 & 2.14 & 5888.02 & 16.30 & 5911.28 & 2.09 & 5945.12 & 10.26  \\
5038.98 & 2.37 & 5698.94 & 4.39 & 5888.45 & 12.91 & 5911.44 & 2.51 & 5945.54 & 8.39  \\
5039.54 & 2.06 & 5700.09 & 4.90 & 5888.65 & 2.56 & 5911.56 & 10.93 & 5946.08 & 4.62  \\
5042.08 & 2.22 & 5701.49 & 5.53 & 5889.51 & 8.52 & 5912.81 & 2.14 & 5946.10 & 4.75  \\
5042.58 & 3.01 & 5712.15 & 2.03 & 5890.43 & 14.86 & 5913.49 & 3.45 & 5946.45 & 8.47  \\
5042.80 & 2.32 & 5718.27 & 3.49 & 5890.88 & 13.39 & 5913.79 & 11.13 & 5946.81 & 20.56  \\
5043.59 & 3.04 & 5719.71 & 2.44 & 5891.00 & 3.11 & 5914.99 & 8.47 & 5947.65 & 10.54  \\
5043.77 & 4.25 & 5720.34 & 4.84 & 5891.94 & 3.19 & 5915.73 & 5.90 & 5947.86 & 18.67  \\
5044.64 & 4.07 & 5722.70 & 2.87 & 5892.28 & 4.57 & 5916.23 & 7.33 & 5949.01 & 2.40  \\
5053.06 & 2.57 & 5722.73 & 2.97 & 5892.45 & 19.40 & 5916.42 & 6.26 & 5949.03 & 2.46  \\
5053.08 & 2.39 & 5724.66 & 2.40 & 5893.19 & 14.27 & 5917.37 & 3.19 & 5949.56 & 2.89  \\
5057.09 & 2.85 & 5727.64 & 3.76 & 5893.84 & 7.46 & 5917.39 & 3.15 & 5949.98 & 17.82  \\
5057.12 & 3.81 & 5730.44 & 3.43 & 5894.30 & 7.48 & 5918.19 & 3.63 & 5950.47 & 2.24  \\
5058.27 & 5.78 & 5736.33 & 2.47 & 5895.18 & 4.93 & 5919.22 & 15.37 & 5950.62 & 7.34  \\
5058.91 & 2.37 & 5738.45 & 5.32 & 5895.75 & 4.22 & 5919.85 & 20.91 & 5950.95 & 2.36  \\
5058.93 & 2.31 & 5742.98 & 3.02 & 5895.91 & 3.10 & 5920.44 & 26.24 & 5951.14 & 5.91  \\
5060.69 & 3.91 & 5746.56 & 4.47 & 5895.93 & 2.95 & 5921.36 & 7.89 & 5951.77 & 2.48  \\
5066.59 & 2.70 & 5754.98 & 2.40 & 5897.28 & 7.47 & 5921.96 & 2.34 & 5952.30 & 6.96  \\
5066.63 & 4.65 & 5755.02 & 2.11 & 5897.63 & 10.41 & 5923.16 & 3.20 & 5955.76 & 10.54  \\
5067.75 & 2.29 & 5762.38 & 2.05 & 5898.25 & 5.24 & 5923.31 & 11.64 & 5957.16 & 5.84  \\
5067.88 & 4.17 & 5764.17 & 2.14 & 5898.73 & 2.66 & 5923.51 & 3.36 & 5957.58 & 4.12  \\
5072.91 & 2.00 & 5860.39 & 2.33 & 5898.95 & 19.89 & 5923.98 & 2.47 & 5958.69 & 13.98  \\
5073.56 & 5.12 & 5860.43 & 2.10 & 5899.80 & 10.22 & 5924.44 & 10.02 & 5959.06 & 8.93  \\
5077.15 & 4.20 & 5871.43 & 3.98 & 5900.72 & 9.82 & 5924.63 & 13.14 & 5959.44 & 12.56  \\
5078.13 & 2.45 & 5872.08 & 2.25 & 5900.83 & 20.36 & 5925.07 & 18.02 & 5960.79 & 3.60  \\
5079.22 & 2.05 & 5876.38 & 3.45 & 5901.71 & 2.94 & 5925.79 & 10.78 & 5962.25 & 2.89  \\
5079.76 & 2.04 & 5876.91 & 5.43 & 5902.03 & 3.98 & 5929.09 & 11.76 & 5963.27 & 5.90  \\
5079.80 & 3.96 & 5877.24 & 4.04 & 5902.26 & 26.44 & 5929.92 & 2.80 & 5965.74 & 2.28  \\
5084.49 & 2.66 & 5880.39 & 4.82 & 5902.93 & 6.63 & 5931.41 & 2.10 & 5965.76 & 2.28  \\
5087.03 & 2.12 & 5880.51 & 5.27 & 5904.33 & 6.40 & 5932.89 & 18.87 & 5967.14 & 3.86  \\
5087.29 & 2.23 & 5881.51 & 3.07 & 5904.65 & 2.29 & 5933.58 & 11.01 & 5967.47 & 10.04  \\
5087.32 & 2.68 & 5881.54 & 2.92 & 5904.67 & 2.32 & 5934.88 & 2.63 & 5968.12 & 2.21  \\
5413.83 & 2.23 & 5881.73 & 5.29 & 5905.91 & 2.62 & 5935.99 & 2.20 & 5968.47 & 2.49  \\
5415.09 & 3.20 & 5881.89 & 4.12 & 5906.09 & 4.50 & 5936.61 & 6.16 & 5968.64 & 11.56  \\
5423.68 & 2.39 & 5882.66 & 5.73 & 5906.12 & 3.91 & 5938.85 & 4.90 & 5968.85 & 2.11  \\
5436.32 & 2.32 & 5882.77 & 4.44 & 5906.15 & 3.09 & 5940.76 & 3.37 & 5969.08 & 12.83  \\            \bottomrule
        \end{tabular}
    \end{threeparttable}
\end{table*}
\begin{table*}
    \begin{threeparttable}
        \centering
        \caption[]{
            \textit{(continued)} List of telluric lines used in this work.
        }
        \label{tab:actlines}
\begin{tabular}{p{1cm}p{1.5cm}|p{1cm}p{1.5cm}|p{1cm}p{1.5cm}|p{1cm}p{1.5cm}|p{1cm}p{1.5cm}}
            \toprule
$\lambda$, \AA & Depth, \% & $\lambda$, \AA & Depth, \% & $\lambda$, \AA & Depth, \% & $\lambda$, \AA & Depth, \% & $\lambda$, \AA & Depth, \%  \\
            \midrule \raggedright
5969.20 & 2.46 & 6282.80 & 25.81 & 6347.85 & 3.21 & 6481.10 & 2.08 & 6534.90 & 9.87  \\
5969.84 & 5.77 & 6283.35 & 3.82 & 6347.89 & 2.63 & 6482.53 & 4.12 & 6535.14 & 2.38  \\
5970.86 & 7.39 & 6283.57 & 20.67 & 6351.56 & 3.32 & 6484.11 & 22.67 & 6535.53 & 2.92  \\
5972.15 & 11.00 & 6284.64 & 16.35 & 6351.58 & 3.13 & 6484.55 & 2.32 & 6535.88 & 2.40  \\
5975.90 & 9.57 & 6285.38 & 8.22 & 6362.10 & 3.23 & 6484.63 & 7.53 & 6537.60 & 6.32  \\
5975.92 & 9.72 & 6286.65 & 6.62 & 6372.44 & 2.11 & 6484.79 & 11.07 & 6541.32 & 2.50  \\
5975.99 & 5.61 & 6288.60 & 14.60 & 6425.79 & 2.89 & 6485.53 & 4.52 & 6543.19 & 9.72  \\
5977.33 & 3.40 & 6289.99 & 2.31 & 6433.82 & 2.64 & 6486.43 & 5.92 & 6544.79 & 26.16  \\
5977.84 & 10.28 & 6290.25 & 18.17 & 6435.42 & 2.47 & 6486.53 & 2.53 & 6544.93 & 2.21  \\
5978.26 & 2.46 & 6291.07 & 24.21 & 6448.79 & 2.82 & 6487.65 & 12.62 & 6546.65 & 4.07  \\
5978.60 & 6.42 & 6291.69 & 2.29 & 6455.02 & 4.12 & 6488.42 & 2.31 & 6548.58 & 11.76  \\
5980.96 & 2.63 & 6293.01 & 25.49 & 6459.74 & 8.11 & 6488.90 & 4.21 & 6549.50 & 19.84  \\
5982.02 & 3.00 & 6293.46 & 5.67 & 6460.55 & 4.15 & 6490.01 & 8.24 & 6552.93 & 2.76  \\
5986.03 & 6.94 & 6293.81 & 30.05 & 6460.86 & 2.41 & 6491.66 & 25.31 & 6553.51 & 22.25  \\
5987.82 & 3.21 & 6295.48 & 3.33 & 6461.08 & 2.20 & 6491.82 & 2.18 & 6553.68 & 2.08  \\
5987.84 & 3.36 & 6295.51 & 3.44 & 6462.69 & 3.08 & 6493.79 & 17.40 & 6554.67 & 8.21  \\
5989.36 & 4.37 & 6296.03 & 28.03 & 6462.71 & 3.80 & 6494.11 & 6.00 & 6558.06 & 13.54  \\
5990.10 & 5.44 & 6296.81 & 32.24 & 6462.73 & 3.77 & 6495.29 & 4.75 & 6559.03 & 6.21  \\
5991.42 & 2.58 & 6298.11 & 5.95 & 6462.86 & 2.72 & 6496.74 & 25.31 & 6561.38 & 6.51  \\
5991.66 & 6.58 & 6299.31 & 28.31 & 6462.89 & 2.82 & 6498.37 & 2.97 & 6561.98 & 4.64  \\
5992.81 & 6.28 & 6300.08 & 31.46 & 6464.35 & 6.00 & 6498.48 & 6.83 & 6563.33 & 2.50  \\
5995.33 & 4.04 & 6302.85 & 25.46 & 6464.65 & 2.05 & 6503.46 & 2.19 & 6564.41 & 6.20  \\
5998.16 & 4.89 & 6303.61 & 27.68 & 6465.29 & 8.10 & 6504.44 & 2.16 & 6564.94 & 2.47  \\
6000.52 & 5.18 & 6306.66 & 20.74 & 6467.00 & 2.75 & 6504.50 & 2.87 & 6564.98 & 2.30  \\
6003.46 & 2.10 & 6307.42 & 22.70 & 6467.14 & 5.23 & 6505.08 & 6.41 & 6565.09 & 14.18  \\
6005.46 & 2.15 & 6310.26 & 2.03 & 6467.59 & 4.57 & 6508.54 & 2.46 & 6569.65 & 2.20  \\
6268.42 & 2.36 & 6310.74 & 15.90 & 6468.44 & 5.69 & 6509.48 & 11.24 & 6569.73 & 4.01  \\
6268.46 & 2.73 & 6311.49 & 16.43 & 6468.75 & 4.17 & 6512.88 & 8.13 & 6572.96 & 14.52  \\
6273.21 & 2.68 & 6315.09 & 10.95 & 6468.82 & 3.20 & 6513.12 & 6.84 & 6575.74 & 22.88  \\
6273.30 & 2.27 & 6315.83 & 11.25 & 6468.84 & 3.08 & 6514.50 & 3.43 & 6580.04 & 2.02  \\
6276.12 & 2.92 & 6317.17 & 5.04 & 6470.23 & 5.43 & 6515.18 & 4.25 & 6581.68 & 7.47  \\
6277.45 & 23.04 & 6318.06 & 4.03 & 6470.51 & 9.69 & 6515.61 & 27.47 & 6584.17 & 2.70  \\
6277.67 & 17.10 & 6318.28 & 2.35 & 6470.87 & 6.62 & 6516.73 & 10.05 & 6584.40 & 3.23  \\
6277.79 & 8.17 & 6319.70 & 7.36 & 6473.34 & 11.43 & 6517.43 & 23.06 & 6584.43 & 3.53  \\
6278.16 & 9.44 & 6320.45 & 6.99 & 6473.48 & 4.90 & 6517.51 & 25.28 & 6586.64 & 2.59  \\
6278.29 & 18.91 & 6320.72 & 2.60 & 6474.05 & 13.43 & 6517.97 & 13.58 & 6587.39 & 8.50  \\
6278.48 & 16.95 & 6320.79 & 4.74 & 6475.00 & 2.52 & 6518.89 & 10.49 & 6587.56 & 4.43  \\
6278.63 & 7.28 & 6322.20 & 2.38 & 6475.92 & 13.86 & 6520.03 & 3.43 & 6587.59 & 3.92  \\
6278.94 & 41.44 & 6324.61 & 6.26 & 6476.07 & 18.40 & 6520.34 & 14.63 & 6595.27 & 4.04  \\
6279.22 & 6.74 & 6324.71 & 3.19 & 6476.68 & 26.54 & 6523.09 & 3.43 & 6600.21 & 6.25  \\
6279.73 & 25.97 & 6324.75 & 3.00 & 6477.88 & 6.04 & 6523.12 & 2.27 & 6603.03 & 3.16  \\
6279.95 & 31.98 & 6325.34 & 3.86 & 6478.20 & 3.65 & 6524.19 & 4.37 & 6606.44 & 2.44  \\
6280.09 & 3.18 & 6329.77 & 2.27 & 6478.36 & 3.04 & 6524.51 & 2.92 & 6606.46 & 2.35  \\
6280.74 & 28.64 & 6330.49 & 2.34 & 6480.06 & 8.64 & 6524.72 & 17.14 & 6613.35 & 2.01  \\
6281.24 & 32.60 & 6332.92 & 3.42 & 6480.35 & 4.81 & 6531.47 & 4.89 & 6613.44 & 3.36  \\
6282.03 & 25.97 & 6343.26 & 4.27 & 6480.93 & 23.10 & 6533.23 & 19.55 & 6835.43 & 2.22  \\        \bottomrule
    \end{tabular}

    \end{threeparttable}
\end{table*}



\bsp    
\label{lastpage}
\end{document}